\documentclass[3p,12pt]{elsarticle}
\biboptions{sort&compress}
\usepackage{lineno,hyperref,isotope,mathrsfs}
\modulolinenumbers[5]
\usepackage{amsmath,bbold,amssymb,epsfig,bm,mathrsfs,feynmp,color,slashed,nicefrac,exscale}
\usepackage[T1]{fontenc}
\usepackage[utf8]{inputenc}
\usepackage{times,txfonts}
\usepackage{isotope}
\usepackage{xcolor}
\usepackage{graphicx}
\usepackage{gensymb}
\usepackage{tikz}
\newcommand{\be}{\begin{equation}}
\newcommand{\ee}{\end{equation}}
\newcommand{\bea}{\begin{eqnarray}}
\newcommand{\eea}{\end{eqnarray}}

\newcommand{\vecq}{{\bm q}}
\newcommand{\veck}{{\bm k}}
\newcommand{\gammavec}{{\bm \gamma}}
\newcommand{\avec}{{\bm \alpha}}
\newcommand{\Tr}{{\rm Tr}}
\newcommand{\ep}{{\varepsilon}}

\newcommand{\prc}{{Phys. Rev. C}}

\def\prc{Phys. Rev. C\ }%
\def\prd{Phys. Rev. D \ }%
\def\physrep{Phys. Rep. \ }%

\definecolor{red}{rgb}{0.8,0,0}
\definecolor{violet}{rgb}{0.4,0,0.4}
\definecolor{green}{rgb}{0,0.5,0.0}
\definecolor{navy}{rgb}{0.0,0.0,0.6}
\definecolor{orange}{rgb}{0.8,0.2,0.0}

\usepackage[normalem]{ulem}  

\begin{document}
\begin{frontmatter}

\title{Color superconductivity from the chiral  quark-meson model}

\author[a]{Armen Sedrakian} 
\ead{sedrakian@fias.uni-frankfurt.de } 
\author[b]{Ralf-Arno Tripolt}
\ead{tripolt@ectstar.eu} 
\author[b,c]{Jochen Wambach}
\ead{jwambach@ectstar.eu} 

\address[a]{ Frankfurt Institute for Advanced Studies, 
Ruth-Moufang-str. 1,  D-60438   Frankfurt-Main, Germany}
\address[b]{European Centre for Theoretical Studies in Nuclear 
  Physics and related Areas (ECT*) and\\ Fondazione 
Bruno Kessler, Villa Tambosi, Strade delle Tabarelle 286, I-38123 
Villazzano (TN), Italy } 
\address[c]{Theoriezentrum, Institut f\"ur Kernphysik, Technische
  Universit\"at Darmstadt, \\Schlossgarten-str. 2, 64289 Darmstadt, Germany}

\begin{abstract}
  We study the two-flavor color superconductivity of low-temperature
  quark matter in the vicinity of chiral phase transition in the
  quark-meson model where the interactions between quarks are
  generated by pion and sigma exchanges.  Starting from the
  Nambu-Gor'kov propagator in real-time formulation we obtain finite
  temperature (real axis) Eliashberg-type equations for the quark
  self-energies (gap functions) in terms of the in-medium spectral
  function of mesons.  Exact numerical solutions of the coupled
  nonlinear integral equations for the real and imaginary parts of the
  gap function are obtained in the zero temperature limit using a
  model input spectral function. We find that these components of the
  gap display a complicated structure with the real part being
  strongly suppressed above $2\Delta_0$, where $\Delta_0$ is its
  on-shell value. We find  $\Delta_0\simeq 40$ MeV close to the chiral
  phase transition.
\end{abstract}


\end{frontmatter}

\section{Introduction}
\label{sec:intro}

Low-temperature quark matter at large chemical potential is expected
to be a color
superconductor~\cite{2008RvMP...80.1455A,2014RvMP...86..509A}. In its
ground state, it forms a coherent state of bound Cooper pairs which
flow without resistivity. At moderate densities, the most robust
pairing pattern involves two light flavors of up and down quarks
forming Cooper pairs with a wave-function that is antisymmetric in
color space~\cite{1984PhR...107..325B}.

Experimental programs exploring highly compressed matter in heavy-ion
collisions will probe the region of the phase diagram of strong
interaction matter where the interplay between the chiral symmetry
breaking and color superconductivity is an important factor
\cite{Friman:2011zz}. In this
regime of interest, which is close to the chiral phase transition
line, quarks and mesons are the dominant degrees of freedom. Having
this context in mind, we address here the 2SC pairing in quark matter
in the quark-meson model, which is a renormalizable model that
shares the chiral symmetry breaking pattern with the underlying
fundamental theory of QCD~\cite{PhysRevD.53.5142,Schaefer2005479}.
More specifically, our work is further motivated by the recent
observation that the entropy of this model shows anomalies at
low-temperatures, when studied within the functional renormalization
group formalism~\cite{Tripolt:2017zgc}. This could be an indication of
the instability of the obtained ground state toward color
superconductivity
or some other phase of QCD, for example, the quarkyonic phase~\cite{2007NuPhA.796...83M}.

Color superconductivity in the 2SC phase was studied at asymptotically
high densities within perturbative QCD framework in
Refs.~\cite{Schaefer_1999,Pisarski_2000,2000PhRvD..61e1501P,2000PhRvD..61e6001H}.
In these theories, the interaction between quarks is mediated via
(screened) gluon exchanges and the pairing fields are governed by
Eliashberg-type equations, familiar from boson-exchange models of
superconductivity. Approximate solutions of these equations for the
case of massless quarks were obtained which exhibit the scaling of the
gap (more precisely, its on-shell value $\Delta_0$) with the strong
coupling $\lambda$ as $\Delta_0\propto \exp (-1/\lambda)$; these
solutions also identified the pre-factor of the (approximate) gap
equation for the real part of the pairing field. However, to our
knowledge, the effects of retardation of interaction via gluon or
other exchanges and the resulting complex nature of the gap function
have not been exposed so far.

The aim of this work is thus to address again the problem of 2SC
pairing, however within a model which is better suited in the regime
close to the chiral phase transition and to maintain the complex
nature of the gap throughout the calculation. We choose to work with
the quark-meson model, where the interaction between quarks is
mediated by pseudo-scalar pion exchanges and scalar sigma
exchanges. The quarks are assumed to be massive due to the dynamical
mechanism of chiral symmetry breaking.  We find the equations for the
2SC pairing gap appropriate for the quark-meson model, which 
naturally encapsulate the information on the spectral functions of
mesons. Furthermore, using an approximate form of the input spectral
functions of mesons we solve the obtained Eliashberg-type equations
exactly, thus fully exhibiting the complex nature of the pairing gap.
 
After this work was completed, Ref.~\cite{2017arXiv171202407A}
appeared which studies pairing in the Yukawa model with a finite-range
interaction and obtains the full energy-momentum dependence of the gap
in the case of imbalanced fermions.  It shows that the frequency
dependence of the gap in the color-flavor-locked phase of QCD has
important ramifications for its color neutrality.

This paper is organized as follows. In Sec.~\ref{sec:formalism} we set
up the formalism for 2SC pairing with the quark-meson model and obtain
the relevant equations for the pairing gap.
Section~\ref{sec:numerical_results} describes the results of numerical
solutions of the gap equations. Our results are
summarized in Sec.~\ref{sec:conclusions}.

\section{Formalism}
\label{sec:formalism}

In this work we apply the Nambu-Gorkov formalism where the quark
states are combined in spinors (our notations follow Ref.~\cite{Peskin:1995ev})
\begin{displaymath}
\label{eq:1.1}
\Psi \equiv \left( \begin{array}{c} \psi \\ {\bar\psi}^T \end{array} 
\right)
             \equiv\left( \begin{array}{c} \psi \\ \psi_c \end{array} 
\right).
\end{displaymath}
The inverse quark propagator, defined in a standard fashion via  
the  Nambu-Gorkov spinors $\Psi$,  is given by 
\be\label{eq:1.2}
S^{-1}(q)
= \left(\begin{array}{cc}{\slashed q}+{\mu}\gamma_0-m&\bar \Delta 
\\
                                 \Delta      & ({\slashed
q}-{\mu}\gamma_0+m)^T \end{array} \right),
\ee
where the following relation holds
$\bar\Delta =\gamma_0\Delta^{\dag}\gamma_0$. We consider the case of
equal number densities of up and down quarks with a common chemical
potential $\mu$ and mass $m$. The real time-structure of the
propagators and self-energies are not specified for simplicity until later.
Furthermore, the vertex corrections to the quark-meson vertices
$ \Gamma_\pi^i(q)$ and $\Gamma_\sigma(q)$ will be neglected 
and these will be approximated by their bare values
\bea
\label{eq:1.3b}
\Gamma_\pi^i(q)=  \left( \begin{array}{cc}\frac{\tau^i}{2}\gamma_5 &0 \\
                 0  &-(\frac{\tau^i}{2}\gamma_5)^T 
\end{array} \right),\quad
\label{eq:1.3c}
\Gamma_\sigma(q)= \left( \begin{array}{cc}\mathbb{I} &0 \\
                 0  &-\mathbb{I}
\end{array} \right),
\eea
where pions are assumed to couple via pseudo-scalar coupling and
$\mathbb{I}$ is a unit matrix in the Dirac and isospin spaces.  The pion and
sigma propagators are given by
\bea\label{eq:1.6}
D_{\pi}(q)=\frac{1}{q_0^2-\vecq^2-m_\pi^2}, \qquad
D_{\sigma}(q)=\frac{1}{q_0^2-\vecq^2-m_\sigma^2},
\eea
where $m_{\pi/\sigma}$ are their masses. 
The gap equation for $\Delta$  in the Fock approximation is then given
by 
\bea\label{eq:1.7}
\Delta(k)
          &=& ig_\pi^2 \int\frac{d^4 q}{(2\pi)^4}
             \left(-\frac{\tau^i}{2}\gamma_5\right)^T S_{21}(q) 
             \frac{\tau^j}{2}\gamma_5\delta_{ij} D_\pi \nonumber\\
&+& ig_\sigma^2 \int\frac{d^4 q}{(2\pi)^4} (-\mathbb{I})^T 
S_{21}(q) \mathbb{I} D_\sigma(q-k) ,
\eea
where $g_\pi$ and $g_\sigma$ are the coupling constants.  The Ansatz
for the gap in a 2SC superconductor is given by~\cite{1984PhR...107..325B}
\bea\label{eq:1.8}
\Delta^{ab}_{ij}(k)=(\lambda_2)^{ab}(\tau_2)_{ij}C\gamma_5
                    [\Delta_+(k)\Lambda^+(k)
                     +\Delta_-(k)\Lambda^-(k)], \quad 
\eea
where $a,b\dots$ refer to the color space, $i,j,\dots$ refer to the
flavor space and the projectors onto the positive and negative states are defined
as $\Lambda^{\pm}(k) =$
$ (E_k^{\pm} + \avec \cdot \veck + m\gamma_0) /2 E_k^{\pm}, $ where
$E_k^{\pm} = \pm\sqrt{\veck^2+m^2}$ and $\avec =
\gamma_0\gammavec$.
Inverting Eq.~\eqref{eq:1.2} one finds for the off-diagonal 21
component of the quark propagator
\bea\label{eq:1.9}
S_{21}(q)=-(\lambda_2\tau_2C\gamma_5)
          \left[\frac{\Delta_+ \Lambda_{-}(q)}{q_0^2-(\epsilon_q-\mu)^2-\Delta_+^2} 
           +\frac{\Delta_-
             \Lambda_{+}(q)}{q_0^2-(\epsilon_q+\mu)^2-\Delta_-^2}\right]
= -(\lambda_2\tau_2C\gamma_5) F_{21}(q).
\eea
On substituting Eqs.~\eqref{eq:1.8} and \eqref{eq:1.9} into
Eq.~\eqref{eq:1.7} and cancelling common terms we find
\bea\label{eq:1.19}  \Delta_+(k)\Lambda^+(k)
+\Delta_-(k)\Lambda^-(k)&=& - ig_\pi^2\frac{3}{4}
\int\frac{d^4 q}{(2\pi)^4}
\gamma_5 F_{21}(q)             \gamma_5 D_\pi(q-k) 
\nonumber\\
&&+ ig_\sigma^2 \int\frac{d^4 q}{(2\pi)^4} F_{21}(q) D_\sigma(q-k).
\eea 
In the next step we decompose the remainder of the anomalous
propagator into a sum of  positive and negative state contributions
 $ F_{21} = \Lambda^- f_1 + \Lambda^+ f_2.$ Now, on multiply
\eqref{eq:1.19} from the right by $\Lambda^+(k)$ and $\Lambda^-(k)$,
using the properties
$(\Lambda^{\pm})^2 = \Lambda^{\pm},$ $\Lambda^+ + \Lambda^- =
1,$ $\Lambda^+\Lambda^- = 0,$
and taking the trace of the resulting two equations (note that
$\Tr~ \Lambda^{\pm} = 4$) we obtain two gap equations introduced 
in Eq.~\eqref{eq:1.8}
\bea\label{eq:1.22a}
  \Delta_+(k)  &=& -\frac{3 ig_\pi^2}{4}
\int\frac{d^4 q}{(2\pi)^4}
            (K_{-+} f_1 +
            K_{++}f_2)D_\pi(q-k) \nonumber\\
&+&
 \frac{ig_\sigma^2}{4} \int\frac{d^4 q}{(2\pi)^4} (M_{-+} f_1 +
            M_{++}f_2)
D_\sigma(q-k).\\ 
\label{eq:1.22b}
     \Delta_-(k)&=& - i\frac{3g_\pi^2}{4}
\int\frac{d^4 q}{(2\pi)^4}
            (K_{--} f_1 +
            K_{+-} f_2)D_\pi(q-k) \nonumber\\
 &+& i\frac{g_\sigma^2}{4} \int\frac{d^4 q}{(2\pi)^4} (M_{--} f_1 +
            M_{+-}f_2) D_\sigma(q-k) 
\eea
where
$K_{\pm\pm} =\Tr[\gamma_5\Lambda^{\pm}(q)\gamma_5\Lambda^{\pm}(k)]$
and $ M_{\pm\pm} =\Tr[\Lambda^{\pm}(q)\Lambda^{\pm}(k)]$.  The
commutation property $[\Lambda^{\pm},\gamma^5] = 0$ implies that we
may set in Eqs.~\eqref{eq:1.22a} and \eqref{eq:1.22b}
$K_{\pm\pm} = M_{\pm\pm}$.  A further simplification arises because
one is generally interested in the gap at the Fermi surface of the
particles and it is legitimate to drop the antiparticle component of the
decomposition of the gap function \eqref{eq:1.8} and take
$\Delta_- = 0$. Indeed the integrand of Eq.~\eqref{eq:1.22a} is
  strongly peaked at the Fermi surface, i.e., when $\epsilon_q=\mu$
  due to the pole structure of the anomalous propagator
  \eqref{eq:1.9}. Its antiparticle pole is located at energies
  $2\mu\sim 700$ MeV and, therefore, cannot influence the physics at
  much lower scale $\sim \Delta_+\ll 2\mu$.

We find then
\bea\label{eq:1.22} \Delta_+(k) =
- i\frac{3g_\pi^2}{4} \int\frac{d^4 q}{(2\pi)^4}
K_{-+} f_1(k-q) D_\pi(q) 
+ i\frac{g_\sigma^2}{4}  \int\frac{d^4 q}{(2\pi)^4} K_{-+} f_1(k-q) D_\sigma(q)
, \eea 
where 
$ f_1 =\Delta_+/(q_0^2-\xi_q^2-\Delta_+^2)$
with $\xi_q = \ep_q-\mu$.
At this point we make explicit the finite-temperature content of the
equations above within the Schwinger-Keldysh real-time formalism.  The
retarded component of the gap function can be written in standard
notations~\cite{Botermans,2003PhRvC..68f5805S}
\bea \Delta_+^R(k_0) =
\int_{-\infty}^{\infty}\frac{d\omega}{2\pi}\int_{-\infty}^{\infty}\frac{d\omega'}{2\pi}
\frac{D^>(\omega')F^>(\omega-\omega')-D^<(\omega')F^<(\omega-\omega')}{k_0-\omega+i\delta},
\eea 
where 
\bea &&F^{>,<}(p) = i A(p) f^{>,<}(p), \quad f^<(p) =
n_F(p), \quad
f^>(p) = 1-n_F(p), \\
&& D^{>,<}(q) = i B(q) g^{>,<}(q), \quad g^<(q) = n_B(q),
\quad g^>(q) = 1+n_B(q), \eea
$n_{F/B}(p)$ are the Fermi and Bose distribution functions, $A(p)$ and
$B(p)$ are the fermionic and bosonic spectral functions;  we  have
suppressed in these equations the pion and sigma indices and momentum
variables which will be restored below.  In terms of these functions
Eq.~\eqref{eq:1.22} can be written as
\bea\label{eq:1.23}
\Delta_+^R(k_0,\veck) &=&
- i\frac{3g_\pi^2}{4}  \int\frac{d^3 q}{(2\pi)^3}
\int_{-\infty}^{\infty}d\ep A(\ep,\veck-\vecq)
\int^{\infty}_{0}\!\!\frac{d\omega'}{2\pi} B_\pi(\omega') {J_\pi(k_0,\omega',\ep)}
K_{-+} 
\nonumber\\
&+& i\frac{g_\sigma^2}{4} \int\frac{d^3 q}{(2\pi)^3} \int_{-\infty}^{\infty}d\ep
A(\ep,\veck-\vecq) \int^{\infty}_{0}\!\!\frac{d\omega'}{2\pi} B_\sigma(\omega')
{J_\sigma(k_0,\omega',\ep)}K_{-+} ,
\eea 
where
\bea\label{eq:1.24}
J_{\pi/\sigma}(k_0,\omega,\ep) = \frac{
      n_{B\,\pi/\sigma }(\omega) + n_F(\ep) }{\ep - k_0- \omega-i\delta}+
    \frac{ 1+n_{B\,\pi/\sigma}(\omega)
      - n_F(\ep) }{\ep -k_0 +\omega-i\delta}
\simeq 
\frac{\theta(-\ep) }{\ep - k_0- \omega-i\delta}+ 
\frac{\theta(+\ep) }{\ep -k_0 +\omega-i\delta} ,
\eea
and the second approximate relation follows in the zero temperature
limit to be used below.  (From now on we drop the sub- and
superscripts on $\Delta$ as we refer only to its retarded, positive 
energy  component).  In the zero temperature
limit, the $d^3q$ phase space integration can be transformed into an
integration over the magnitude of $q$ and the on-shell energy $\xi_p$,
which can be then performed analytically. As a result we find
\bea\label{eq:1.25}
\Delta(k_0,k_F) = 
\int_{0}^{\infty}d\ep ~ \mathscr{F}(\ep)\int^{\infty}_{0}{d\omega'} \lambda(\omega')
\left[
\frac{1}{\ep + k_0+ \omega'+i\delta}+\frac{ 1}{\ep -k_0 +\omega'-i\delta} 
\right],
\eea
where the kernel of the gap equation is given by 
\bea\label{eq:1.26a}
\lambda (\omega) = \frac{g_\sigma^2 \mathscr{B}_{\sigma}(\omega)
-3g_\pi^2\mathscr{B}_{\pi}(\omega)}{4v_F},
\eea
where $v_F$ is the Fermi velocity of quarks and 
\bea\label{eq:1.26b}
\mathscr{B}_{\pi/\sigma}(\omega) = \int_0^{2k_F}
\frac{qdq}{(2\pi)^2} B_{\pi/\sigma}(\omega, q)K_{-+} ,\quad 
\mathscr{ F}(\ep) = {\rm Re}
\frac{\Delta(\ep){\rm sgn}(\ep)}{[\ep^2-\Delta(\ep)^2]^{1/2}}.
\eea
In the case where the spectral function $B_{\pi/\sigma}(\omega, q)$ does not depend
on the momentum transfer $q$, the first equation in
\eqref{eq:1.26b} simplifies to
$\mathscr{B}_{\pi/\sigma}(\omega) \simeq ({m^2k_F^2}/{\pi^2E_F^2})
B_{\pi/\sigma}(\omega)$,
where we substituted the zero temperature limit of $K_{-+}(q\to 0)$. 
Then, the kernel can be written as 
\bea\label{eq:1.27}
\lambda (\omega) = (m^2 v_F/4\pi^2) \left[g_\sigma^2 {B}_{\sigma}(\omega)
  -3g_\pi^2{B}_{\pi}(\omega)\right],
\eea
i.e., up to a constant factor, it is given by the sum of the 
spectral functions of mesons.

\section{Numerical results}
\label{sec:numerical_results}

\begin{figure}[t] 
\begin{center}
\includegraphics[width=9cm,height=7cm,angle=0]{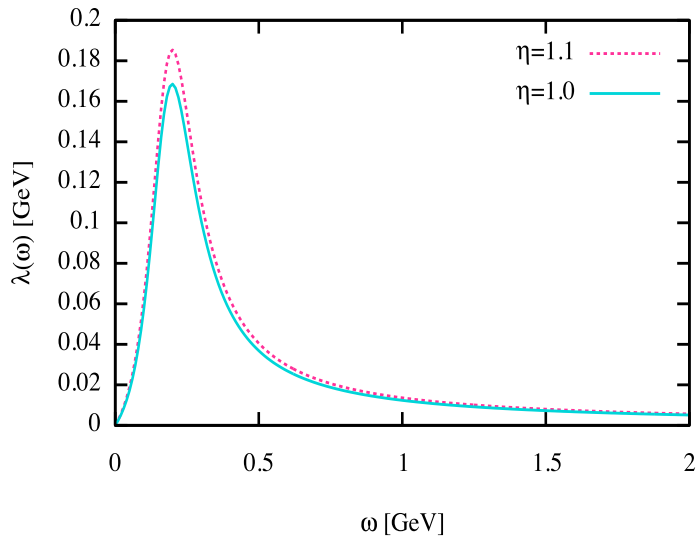}
\caption{ Dependence of the kernel function given by \eqref{eq:2.1}
  for two values of the strengths of interaction parametrized by
  $\eta$ on frequency.}
\label{fig:1}
\end{center}
\end{figure}

Eq.~\eqref{eq:1.25} represents two coupled non-linear integral equations
for the real and imaginary parts of the gap function, which were solved
iteratively on a quadratic mesh spanned by the variables
$[\epsilon, \omega]$. The numerical method has been described
elsewhere~\cite{2003PhRvC..68f5805S}.  We approximate the kernel of
the gap function, Eq.~\eqref{eq:1.27}, by a suitable Gaussian function
of the form
\bea\label{eq:2.1}
\lambda (\omega)  
= \frac{g\omega}{(\omega-\omega_0)^2 + \gamma^2/4},
\eea
with the parameter values chosen as $\gamma = 0.0972$, $g =0.0077$ and
$\omega_0 = 0.1734.$ To obtain these parameter values we have computed
Eq.~\eqref{eq:1.27} using as an input the spectral functions
$B_{\pi/\sigma}(\omega)$ derived from the quark-meson
model~\cite{PhysRevD.89.034010}.  The centroid of Eq.~\eqref{eq:2.1}
$\omega_0$ is at the mass of the $\sigma$ meson and its hight $g$ was
matched to the numerical computation of Eq.~\eqref{eq:1.27}.  To
explore the sensitivity of the result on the strength of the coupling
we repeated the computations by rescaling $g \to \eta g$, where $\eta$
is a constant factor.  In Fig.~\ref{fig:1} we plot the function
$\lambda(\omega)$ in Eq. \eqref{eq:2.1} for two values of $\eta$
indicated in the figure.

\begin{figure}[t] 
\begin{center}
\includegraphics[width=9cm,height=7cm,angle=0]{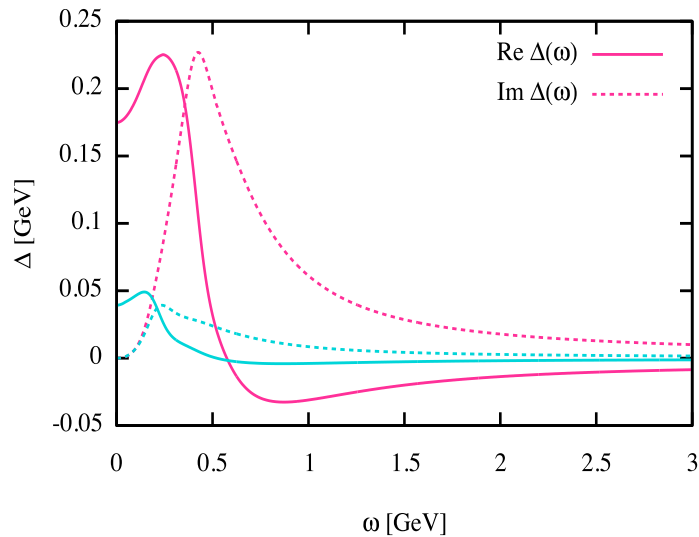}
\caption{ Dependence of the real (solid) and imaginary (dashed)
  components of the gap function on frequency for two value strengths
  of interaction $\eta =1 $ (lower) and $\eta =1.1 $ (upper) pair of
  curves.}
\label{fig:2}
\end{center}
\end{figure}

The solutions of the gap equation are shown in Fig.~\ref{fig:2}, where
we display the real and imaginary parts of the gap as a function of
frequency. The on-shell value of the gap $\Delta_0$ follows in the
limit $\omega = 0$ where it becomes purely real; it is seen that this
value is rather sensitive to the strength parameter $\eta$. Increasing its
value by $10\%$ produces a four-fold increase in
$\Delta_0$. Computations for a larger value $\eta = 1.4$ (not shown in
the figure) display a further increase of the gap value up to
$\Delta_0\simeq 0.3$~GeV.

In the off-shell region, the imaginary and real parts of the gap show
non-trivial structures. They intersect for $\omega \simeq 2\Delta_0$,
beyond which the imaginary component dominates before both components
vanish at asymptotically large frequencies.  Note that in the ordinary
BCS formulations the gap is real and constant in the off-shell
region. Clearly, our results show that the constant gap approximation
could be accurate only very close to the on-shell ($\omega\to 0$)
limit. A proper account of the frequency dependence of the propagators
of the color superconductors may be of importance for many frequency
dependent observables, for example, for the description of their
dynamical response to various perturbations. Examples include the
dynamical (frequency dependent) Meissner effect or transport
coefficients, such as shear viscosity.  We recall that in the
framework of the Kubo formalism, see e.g.~\cite{2017PhRvD..95k4021H},
the last quantity requires an evaluation of the frequency derivative
of response function, which will obtain an additional contribution
through the frequency dependence of the gap function.

\section{Conclusions and perspectives}
\label{sec:conclusions}

We have set up a formalism to compute the pairing gap in the 2SC phase
of low-temperature quark matter within the quark-meson model.
Starting from the Nambu-Gorkov propagator of the quarks for the 2SC
phase we have evaluated their anomalous self-energy (gap) due to meson
exchanges. Using the real-time formalism we have expressed the gap
function in terms of spectral functions of mesons (here pions and
sigmas) at finite temperatures, see Eq.~\eqref{eq:1.23}. The
frequency dependence of the spectral functions implies a complex gap
function, which physically reflects the retardation of the pairing
interaction (which is absent in the BCS-type formulations). We have
solved the coupled integral equations for the real and imaginary parts
of the gap function in the zero-temperature limit, showing that these
components have non-trivial structures in the frequency domain, see
Fig.~\ref{fig:2}.

For the sake of physical insight and simplicity, we have approximated
the full spectral functions of the quark-meson model by a
Gaussian-type function and explored the dependence of the gap on the
strength of the interaction. We find that the on-shell value of the
gap strongly depends on the strengths of the attraction in the pairing
channel, which is consistent with the expectations from the BCS type
approaches. It would be interesting to evaluate the components of the
2SC gap function using spectral functions of the quark-meson model
directly for specific values of density and temperature of quark matter. 

The frequency dependence and complex nature of the gap function
implies that a number of physical quantities may differ qualitatively
from their BCS counterparts computed with a real, constant in the
frequency domain, gap. Among many examples, the transport
coefficients, such as shear
viscosity~\cite{2014PhRvC..90e5205A,
2017PhRvD..96i4025S,2017PhRvD..95k4021H},
would be interesting to evaluate.

\section*{Acknowledgments} 
We thank M. Alford, M. Buballa, A. Harutyunyan and D. Rischke for
discussions.  A.~S. was supported by the Deutsche
Forschungsgemeinschaft (DFG) Grant No. SE 1836/3-2 and thanks ECT$^*$
Trento for its generous hospitality.  The work of R.-A.~T. and
J.~W. was supported by the DFG through the grant CRC-TR 211.


\begin{thebibliography}{20}
\expandafter\ifx\csname natexlab\endcsname\relax\def\natexlab#1{#1}\fi
\providecommand{\url}[1]{\texttt{#1}}
\providecommand{\href}[2]{#2}
\providecommand{\path}[1]{#1}
\providecommand{\DOIprefix}{doi:}
\providecommand{\ArXivprefix}{arXiv:}
\providecommand{\URLprefix}{URL: }
\providecommand{\Pubmedprefix}{pmid:}
\providecommand{\doi}[1]{\href{http://dx.doi.org/#1}{\path{#1}}}
\providecommand{\Pubmed}[1]{\href{pmid:#1}{\path{#1}}}
\providecommand{\bibinfo}[2]{#2}
\ifx\xfnm\relax \def\xfnm[#1]{\unskip,\space#1}\fi
\bibitem[{{Alford} et~al.(2008){Alford}, {Schmitt}, {Rajagopal}, and
  {Sch{\"a}fer}}]{2008RvMP...80.1455A}
\bibinfo{author}{M.~G. {Alford}}, \bibinfo{author}{A.~{Schmitt}},
  \bibinfo{author}{K.~{Rajagopal}}, \bibinfo{author}{T.~{Sch{\"a}fer}},
\newblock \bibinfo{title}{{Color superconductivity in dense quark matter}},
\newblock \bibinfo{journal}{Reviews of Modern Physics} \bibinfo{volume}{80}
  (\bibinfo{year}{2008}) \bibinfo{pages}{1455--1515}.
\bibitem[{{Anglani} et~al.(2014){Anglani}, {Casalbuoni}, {Ciminale},
  {Ippolito}, {Gatto}, {Mannarelli}, and {Ruggieri}}]{2014RvMP...86..509A}
\bibinfo{author}{R.~{Anglani}}, \bibinfo{author}{R.~{Casalbuoni}},
  \bibinfo{author}{M.~{Ciminale}}, \bibinfo{author}{N.~{Ippolito}},
  \bibinfo{author}{R.~{Gatto}}, \bibinfo{author}{M.~{Mannarelli}},
  \bibinfo{author}{M.~{Ruggieri}},
\newblock \bibinfo{title}{{Crystalline color superconductors}},
\newblock \bibinfo{journal}{Reviews of Modern Physics} \bibinfo{volume}{86}
  (\bibinfo{year}{2014}) \bibinfo{pages}{509--561}.
\bibitem[{{Bailin} and {Love}(1984)}]{1984PhR...107..325B}
\bibinfo{author}{D.~{Bailin}}, \bibinfo{author}{A.~{Love}},
\newblock \bibinfo{title}{{Superfluidity and superconductivity in relativistic
  fermion systems}},
\newblock \bibinfo{journal}{\physrep} \bibinfo{volume}{107}
  (\bibinfo{year}{1984}) \bibinfo{pages}{325--385}.
\bibitem[{Friman et~al.(2011)Friman, Hohne, Knoll, Leupold, Randrup, Rapp, and
  Senger}]{Friman:2011zz}
\bibinfo{author}{B.~Friman}, \bibinfo{author}{C.~Hohne},
  \bibinfo{author}{J.~Knoll}, \bibinfo{author}{S.~Leupold},
  \bibinfo{author}{J.~Randrup}, \bibinfo{author}{R.~Rapp},
  \bibinfo{author}{P.~Senger},
\newblock \bibinfo{title}{{The CBM physics book: Compressed baryonic matter in
  laboratory experiments}},
\newblock \bibinfo{journal}{Lect. Notes Phys.} \bibinfo{volume}{814}
  (\bibinfo{year}{2011}) \bibinfo{pages}{pp.1--980}.
\bibitem[{Jungnickel and Wetterich(1996)}]{PhysRevD.53.5142}
\bibinfo{author}{D.-U. Jungnickel}, \bibinfo{author}{C.~Wetterich},
\newblock \bibinfo{title}{Effective action for the chiral quark-meson model},
\newblock \bibinfo{journal}{Phys. Rev. D} \bibinfo{volume}{53}
  (\bibinfo{year}{1996}) \bibinfo{pages}{5142--5175}.
\bibitem[{Sch\"afer and Wambach(2005)}]{Schaefer2005479}
\bibinfo{author}{B.-J. Sch\"afer}, \bibinfo{author}{J.~Wambach},
\newblock \bibinfo{title}{The phase diagram of the quark--meson model},
\newblock \bibinfo{journal}{Nuclear Physics A} \bibinfo{volume}{757}
  (\bibinfo{year}{2005}) \bibinfo{pages}{479 -- 492}.
\bibitem[{Tripolt et~al.(2017)Tripolt, Schäfer, von Smekal, and
  Wambach}]{Tripolt:2017zgc}
\bibinfo{author}{R.-A. Tripolt}, \bibinfo{author}{B.-J. Schäfer},
  \bibinfo{author}{L.~von Smekal}, \bibinfo{author}{J.~Wambach},
\newblock \bibinfo{title}{The low-temperature behavior of the quark-meson
  model},  
\newblock \bibinfo{journal}{Phys. Rev. D} \bibinfo{volume}{97}
  (\bibinfo{year}{2018}) \bibinfo{pages}{034022}.
\bibitem[{{McLerran} and {Pisarski}(2007)}]{2007NuPhA.796...83M}
\bibinfo{author}{L.~{McLerran}}, \bibinfo{author}{R.~D. {Pisarski}},
\newblock \bibinfo{title}{{Phases of dense quarks at large N$_{}$}},
\newblock \bibinfo{journal}{Nuclear Physics A} \bibinfo{volume}{796}
  (\bibinfo{year}{2007}) \bibinfo{pages}{83--100}.
\bibitem[{{Sch{\"a}fer} and {Wilczek}(1999)}]{Schaefer_1999}
\bibinfo{author}{T.~{Sch{\"a}fer}}, \bibinfo{author}{F.~{Wilczek}},
\newblock \bibinfo{title}{{Superconductivity from perturbative one-gluon
  exchange in high density quark matter}},
\newblock \bibinfo{journal}{\prd} \bibinfo{volume}{60} (\bibinfo{year}{1999})
  \bibinfo{pages}{114033}.
\bibitem[{{Pisarski} and {Rischke}(2000{\natexlab{a}})}]{Pisarski_2000}
\bibinfo{author}{R.~D. {Pisarski}}, \bibinfo{author}{D.~H. {Rischke}},
\newblock \bibinfo{title}{{Color superconductivity in weak coupling}},
\newblock \bibinfo{journal}{\prd} \bibinfo{volume}{61}
  (\bibinfo{year}{2000}{\natexlab{a}}) \bibinfo{pages}{074017}.
\bibitem[{{Pisarski} and {Rischke}(2000{\natexlab{b}})}]{2000PhRvD..61e1501P}
\bibinfo{author}{R.~D. {Pisarski}}, \bibinfo{author}{D.~H. {Rischke}},
\newblock \bibinfo{title}{{Gaps and critical temperature for color
  superconductivity}},
\newblock \bibinfo{journal}{\prd} \bibinfo{volume}{61}
  (\bibinfo{year}{2000}{\natexlab{b}}) \bibinfo{pages}{051501}.
\bibitem[{{Hong} et~al.(2000){Hong}, {Miransky}, {Shovkovy}, and
  {Wijewardhana}}]{2000PhRvD..61e6001H}
\bibinfo{author}{D.~K. {Hong}}, \bibinfo{author}{V.~A. {Miransky}},
  \bibinfo{author}{I.~A. {Shovkovy}}, \bibinfo{author}{L.~C.~R.
  {Wijewardhana}},
\newblock \bibinfo{title}{{Schwinger-Dyson approach to color superconductivity
  in dense QCD}},
\newblock \bibinfo{journal}{\prd} \bibinfo{volume}{61} (\bibinfo{year}{2000})
  \bibinfo{pages}{056001}.
\bibitem[{{Alford} et~al.(2017){Alford}, {Pangeni}, and
  {Windisch}}]{2017arXiv171202407A}
\bibinfo{author}{M.~G. {Alford}}, \bibinfo{author}{K.~{Pangeni}},
  \bibinfo{author}{A.~{Windisch}},
\newblock \bibinfo{title}{{Color superconductivity and charge neutrality in
  Yukawa theory}},
\newblock \bibinfo{journal}{Phys. Rev. Lett.} \bibinfo{volume}{120}
  (\bibinfo{year}{2018}) \bibinfo{pages}{082701}.
\bibitem[{Peskin and Schroeder(1995)}]{Peskin:1995ev}
\bibinfo{author}{M.~E. Peskin}, \bibinfo{author}{D.~V. Schroeder},
  \bibinfo{title}{{An Introduction to quantum field theory}},
  \bibinfo{publisher}{Addison-Wesley}, \bibinfo{address}{Reading, USA},
  \bibinfo{year}{1995}.
\bibitem[{{Botermans} and {Malfliet}(1990)}]{Botermans}
\bibinfo{author}{W.~{Botermans}}, \bibinfo{author}{R.~{Malfliet}},
\newblock \bibinfo{title}{{Quantum transport theory of nuclear matter}},
\newblock \bibinfo{journal}{\physrep} \bibinfo{volume}{198}
  (\bibinfo{year}{1990}) \bibinfo{pages}{115--194}.
\bibitem[{{Sedrakian}(2003)}]{2003PhRvC..68f5805S}
\bibinfo{author}{A.~{Sedrakian}},
\newblock \bibinfo{title}{{Off-shell pairing correlations from meson-exchange
  theory of nuclear forces}},
\newblock \bibinfo{journal}{\prc} \bibinfo{volume}{68} (\bibinfo{year}{2003})
  \bibinfo{pages}{065805}.
\bibitem[{Tripolt et~al.(2014)Tripolt, Strodthoff, von Smekal, and
  Wambach}]{PhysRevD.89.034010}
\bibinfo{author}{R.-A. Tripolt}, \bibinfo{author}{N.~Strodthoff},
  \bibinfo{author}{L.~von Smekal}, \bibinfo{author}{J.~Wambach},
\newblock \bibinfo{title}{Spectral functions for the quark-meson model phase
  diagram from the functional renormalization group},
\newblock \bibinfo{journal}{Phys. Rev. D} \bibinfo{volume}{89}
  (\bibinfo{year}{2014}) \bibinfo{pages}{034010}.
\bibitem[{{Harutyunyan} et~al.(2017){Harutyunyan}, {Rischke}, and
  {Sedrakian}}]{2017PhRvD..95k4021H}
\bibinfo{author}{A.~{Harutyunyan}}, \bibinfo{author}{D.~H. {Rischke}},
  \bibinfo{author}{A.~{Sedrakian}},
\newblock \bibinfo{title}{{Transport coefficients of two-flavor quark matter
  from the Kubo formalism}},
\newblock \bibinfo{journal}{\prd} \bibinfo{volume}{95} (\bibinfo{year}{2017})
  \bibinfo{pages}{114021}.
\bibitem[{{Alford} et~al.(2014){Alford}, {Nishimura}, and
  {Sedrakian}}]{2014PhRvC..90e5205A}
\bibinfo{author}{M.~G. {Alford}}, \bibinfo{author}{H.~{Nishimura}},
  \bibinfo{author}{A.~{Sedrakian}},
\newblock \bibinfo{title}{{Transport coefficients of two-flavor superconducting
  quark matter}},
\newblock \bibinfo{journal}{\prc} \bibinfo{volume}{90} (\bibinfo{year}{2014})
  \bibinfo{pages}{055205}.
\bibitem[{{Sarkar} and {Sharma}(2017)}]{2017PhRvD..96i4025S}
\bibinfo{author}{S.~{Sarkar}}, \bibinfo{author}{R.~{Sharma}},
\newblock \bibinfo{title}{{Shear viscosity of two-flavor inhomogenous color
  superconducting quark matter}},
\newblock \bibinfo{journal}{\prd} \bibinfo{volume}{96} (\bibinfo{year}{2017})
  \bibinfo{pages}{094025}.

\end{thebibliography}
\end{document}